\documentclass[aps,prl,twocolumn,amsmath,amssymb,superscriptaddress,floatfix,showpacs]{revtex4}
\usepackage{graphicx}
\usepackage{bm}
\input epsf.sty
\begin{document}
\title{Parameter free scaling relation for nonequilibrium growth processes}
\author{Yen-Liang Chou}
\author{Michel Pleimling}
\affiliation{Department of Physics, Virginia Polytechnic Institute and State University, Blacksburg, Virginia 24061-0435, USA}

\begin{abstract}
We discuss a parameter free scaling relation that yields
a complete data collapse for large classes of nonequilibrium growth processes.
We illustrate the power of this new scaling relation through various growth models, as for example the
competitive growth model
RD/RDSR (random deposition/random deposition with surface diffusion) and the RSOS (restricted solid-on-solid) model
with different nearest-neighbor height differences, as well as through a new deposition
model with temperature dependent diffusion. The new scaling relation is compared to the
familiar Family-Vicsek relation and the limitations of the latter are highlighted. 

\end{abstract}


\maketitle

The study of growing interfaces has been a very active field
for many years \cite{Mea93,Hal95,Bar95}. Many studies focus on the technologically relevant
growth of thin films or nanostructures, but growing interfaces
are also encountered in various other physical, chemical, or biological systems, ranging from bacterial growth to 
diffusion fronts. Over the years important insights into the behavior of nonequilibrium growth processes have been
gained through the study of simple model systems that capture the most important aspects of real experimental 
systems \cite{Kru97,Kru95}.

\begin{figure}[h]
\centerline{\epsfxsize=3.25in\ \epsfbox{
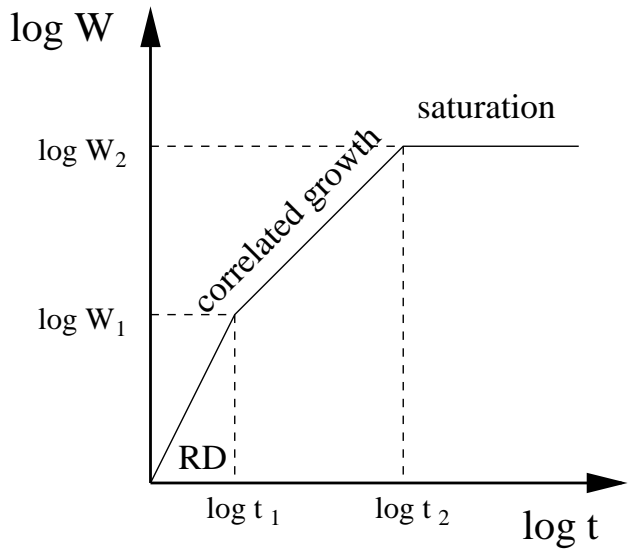}}
\caption{Schematic plot of the interface width as a function of time for a
typical deposition process. The early time behavior (with $t < t_1$) is that of the 
random deposition (RD) process. For $t > t_1$ correlated growth sets in.
Finally, the finite system
displays at late times $t > t_2$ a crossover to a saturation regime where the width of the interface 
remains constant, $W = W_2$.
\label{fig1}
}
\end{figure}

In their seminal work, Edwards and Wilkinson investigated
surface growth phenomena generated by particle sedimentation under the influence of gravity \cite{1982SE}. 
They proposed to describe this process in ($d+1$) dimensions by the following stochastic equation of motion
for the surface height $h(\textbf{x},t)$, now called the Edwards-Wilkinson (EW) equation,
\begin{equation}
\frac{\partial h(\textbf{x},t)}{\partial t}=\nu \nabla^2 h(\textbf{x},t)+\eta(\textbf{x},t),
\label{eqEW}
\end{equation}
where $\nu$ is the diffusion constant (surface tension), whereas $\eta(\textbf{x},t)$ is a Gaussian white noise with zero mean
and covariance $\left< \eta(\textbf{x},t) \eta(\textbf{y},s) \right> = D \delta^d(\textbf{x} - \textbf{y} ) \delta(t-s)$.
Since Eq.~\ref{eqEW} is linear, it can be solved exactly by Fourier transformations \cite{1982SE,Hal95,Kru97}. 
Later, Family \cite{1986FF} discussed the random deposition (RD) and 
random deposition with surface relaxation (RDSR) processes.
RD \cite{1986FF,Bar95} is one 
of the simplest surface growth processes. In this lattice model particles drop from randomly chosen sites over the surface and 
stick directly on the top of the selected surface site. Since there is no surface diffusion, the independently growing columns 
yield an uncorrelated and never-saturated surface. The RDSR process is realized by adding surface diffusion which allows
particles just deposited on the surface to jump to the neighboring site with lowest height.
This diffusion step smoothes the surface and limits the maximum interface width $W(t)$, defined at deposition
time $t$ as 
the standard deviation of the surface height $h$ from its mean value $\overline{h}$:
$W(t) = \sqrt{\left< \left( h - \overline{h} \right)^2 \right> }.$ 
Starting from an initially flat surface, RDSR yields at very early times, with $t <  t_1 \sim 1$ 
(we assume here that one layer is deposited per unit time),
a surface growing in the same way
as for the RD process since no (or only very few)
diffusion steps occur in that regime. For $t > t_1$ the width increases as a power law of time 
with a growth exponent $\beta$ before entering a saturation regime after a crossover time $t_2$, see Fig. \ref{fig1}. 
Both the saturation width $W_2$ and the crossover time $t_2$ are powers of the substrate size $L$:
\begin{equation} 
W_2  \sim  L^\alpha ~, ~~~~t_2  \sim  L^z~,\label{eqWs}
\end{equation} 
where $\alpha$ is the roughness exponent and $z$ is the dynamical exponent, with
$z= \alpha/\beta$.
In his study Family noticed that the scaling exponents obtained through numerical simulations of the RDSR process
agree with those obtained from the solution of the EW equation. 
The dependence of the growing interface on the substrate size $L$ is summarized in the celebrated
Family-Vicsek scaling relation \cite{1985FF}
\begin{equation}
W=L^\alpha f(t/L^z)~.
\label{eqFV}
\end{equation}
Combining this with the relations given in (\ref{eqWs}), we see that this scaling relation mainly consists
in shifting the crossover points for the different system sizes to a common point with the new coordinates
$t' = t/L^z \sim t/t_2$ and $W' = W/L^\alpha \sim W/W_2$.
It is worth noting that the Family-Vicsek relation neglects the RD regime at early times and exclusively
focuses on the two regimes connected by the crossover point at $t = t_2$.

The scaling (\ref{eqWs}) and (\ref{eqFV}) is generic for growing interfaces and has been verified analytically,
numerically, and experimentally in a large variety of systems. Various universality classes have been identified
which differ by the values of the scaling exponents. Thus the RDSR process belongs to the Edwards-Wilkinson universality
class with the exponents $\beta = 1/4$ and $z=2$ for a one-dimensional substrate. 
The RD process is in an universality class of its own which for a d=1 substrate is characterized by the values
$\beta = 1/2$ and $z = \infty$. Other well known universality classes, directly related to
technologically relevant growth processes, are the Kardar-Parisi-Zhang (KPZ) \cite{Kar86} and the conserved KPZ universality
classes \cite{Wol90,Sar91}.

\begin{figure}[h]
\centerline{\epsfxsize=3.25in\ \epsfbox{
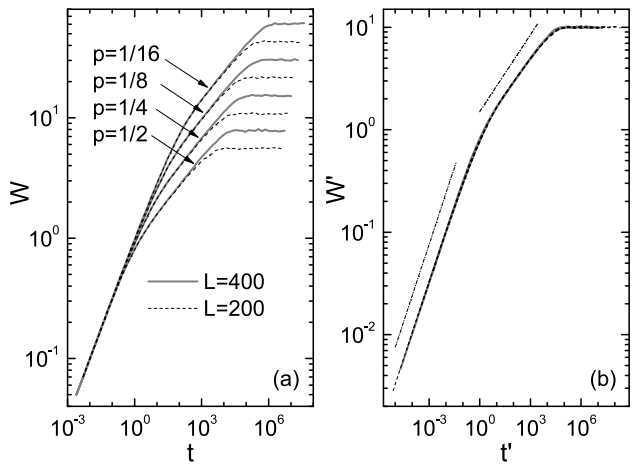}}
\caption{(a) Log-log plot of the surface width vs time for the RD/RDSR process in
systems with different sizes $L$ and different probabilities $p$.
(b) A complete data collapse of all data sets is achieved when using the scaling relation (\ref{eqscal}).
The curves shown in the left panel completely fall one on top of the other and are no longer distinguishable. The axis labels
are $t' = t^{\lambda}/t_1^{\lambda}$ and $W'= W^{\lambda}/W_1^{\lambda}$, see main text. 
The dashed lines indicate the expected
slopes in the RD and EW regimes.
\label{fig2}
}
\end{figure}

\begin{figure}[h]
\centerline{\epsfxsize=3.25in\ \epsfbox{
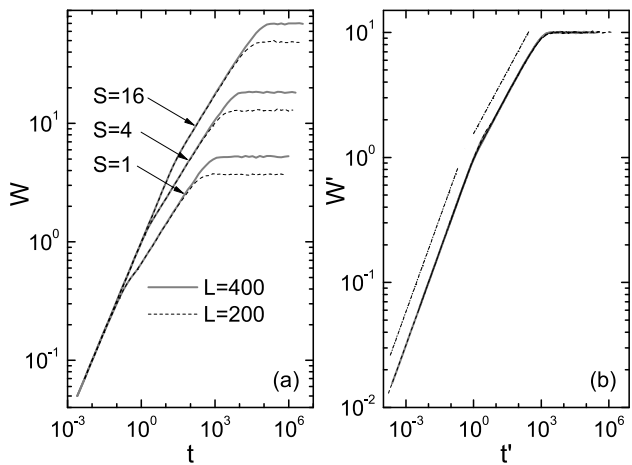}}
\caption{The same as Fig. \ref{fig2}, but now for the RSOS model with different values $S$ of the
maximal nearest-neighbor height difference. Again a perfect data collapse is achieved
when using the scaling relation (\ref{eqscal}).
The dashed lines indicate the expected
slopes in the RD and KPZ regimes.
\label{fig3}
}
\end{figure}

In recent years the study of nonequilibrium growth systems has shifted to more complex cases
as for example competitive growth models, see e.g.
\cite{Hor01,Hor01a,Cha02,Hor03,Kol04,Mur04,Bra05,Iru05,Hor06,Rei06,Oli06,Kol06}.
In a competitive growth model one considers
a mixture of two different deposition processes where one of them takes place with probability $p$ whereas the
other takes place with probability $1-p$. One example is the RD/RDSR model \cite{Hor01} where the deposition happens
according to the RDSR rules with probability $p$ and to the RD rules with probability $1-p$.
Whereas for $p=1$ and $p=0$ only one of the processes is realized,
for general values of $p$ the mixture of the two processes leads to a crossover between the two regimes
where the crossover time and width depend on the value of $p$ (see Fig.\ \ref{fig2}a). A similar dependence on system parameters is
also observed in the RSOS model \cite{Kim89} which exhibits a crossover from the RD regime to the important KPZ universality class.
In this model new particles are incorporated into the growing surface only if the height differences between the deposition site and
its neighboring sites remain smaller than some maximum height $S$. As discussed in \cite{Chi04} and shown in 
Fig. \ref{fig3}a, the crossover time and width depend
on the value of $S$.


In simple growth processes the random deposition regime is restricted to very early times.
This is fundamentally different 
in more complex systems where the initial regime can extend over very large times
\cite{Hor01,Hor01a,Cha02,Hor03,Kol04,Mur04,Bra05,Iru05,Hor06,Rei06,Oli06,Kol06,Chi04}.
As already mentioned, the Family-Vicsek scaling relation (\ref{eqFV}) assigns a new set of
coordinates to the second crossover point.
This does however not yield a complete data collapse
for growth processes with two crossover points if one considers systems of different sizes.
For the competitive growth models some phenomenological
scaling relations have been proposed in the past, but
these modified scaling relations also only allow a partial collapse of
the different curves \cite{Hor01,Kol04,Bra05,Kol06}.

However, a scaling relation leading to a complete data collapse of {\it all} curves obtained for different system sizes and
different values of the system parameters can indeed be obtained for any
growth system that exhibits two different crossover points.
This data collapse is achieved in a two-step process. First
we translate all curves in a log-log plot such that the first crossover point is now located at the origin. This is achieved by plotting $\log \left( \frac{W}{W_1} \right)$ as a function of $\log \left( \frac{t}{t_1} \right)$. In the second step
we rescale both axes by the common scale factor $\lambda = 1/\log \left( \frac{W_2}{W_1} \right)$ such that in the log-log
plot the second crossover point is fixed at the rescaled width $\log W'_2 = 1$. This isotropic rescaling, which conserves
the slope of the region between the two crossover points, makes that the length of the line connecting the two crossover
points is the same for all curves, and a complete data collapse, encompassing all three regimes, follows.
Our proposed scaling relation can be cast in the following equation:
\begin{equation}
\frac{\log \left( \frac{W}{W_1} \right)}{\log \left( \frac{W_2}{W_1} \right)}= F \left[
\frac{\log \left( \frac{t}{t_1} \right)}{\log \left( \frac{W_2}{W_1} \right)} \right]
\end{equation}
where $F(x)$ is a scaling function. Introducing $\lambda = 1/\log \left( \frac{W_2}{W_1} \right)$, we can rewrite
this as
\begin{equation} \label{eqscal}
W^\lambda = W_1^\lambda \, G \left[ \left( \frac{t}{t_1} \right)^\lambda \right]
\end{equation}
with a new scaling function $G(y)$. As shown in Fig. \ref{fig2} and \ref{fig3} for the
RD/RDSR and RSOS processes,
the proposed scaling relation yields a complete data collapse for different system sizes $L$ and different
values of the system parameters. This perfect scaling
behavior should be compared with the incomplete scaling proposed
in the literature \cite{Hor01,Kol04,Bra05,Kol06}.

Obviously, the scaling relation (\ref{eqscal}) is of universal use in growth systems
with two crossover points and replaces the Family-Vicsek relation in these systems.
This class of systems encompasses competitive growth models, but also 
the simple growth systems, for which the Family-Vicsek relation has been proposed originally,
belong to this class.
It is also worth noting that the properties of the different models only enter in our
relation (\ref{eqscal}) implicitly through the
dependence of the positions of the crossover points on the different system parameters.

We can also state the new scaling relation in an alternative way which makes the difference to the
Family-Vicsek relation more transparent. Indeed, a complete collapse can also be achieved when first moving
the second crossover point to the origin in a log-log plot, yielding the
relation
\begin{equation}
\frac{\log \left( \frac{W}{W_2} \right)}{\log \left( \frac{W_2}{W_1} \right)}= \tilde{F} \left[
\frac{\log \left( \frac{t}{t_2} \right)}{\log \left( \frac{W_2}{W_1} \right)} \right]
\end{equation}
or
\begin{equation} \label{eqscal2}
W^\lambda = W_2^\lambda \, \tilde{G} \left[ \left( \frac{t}{t_2} \right)^\lambda \right]
\end{equation}
where $\tilde{F}$ and $\tilde{G}$ are again scaling functions. This scaling relation is completely
equivalent to the first one, only the scales are shifted. In fact, the relation (\ref{eqscal2})
allows a direct comparison with the Family-Vicsek relation (\ref{eqFV}):
recalling that the scaling behaviors of $W_2$ and $t_2$ are given by the relations (\ref{eqWs}),
we immediately see that we recover the Family-Vicsek relation by setting $\lambda =1$.
This nicely shows that it is the isotropic rescaling in the log-log plot by the factor $\frac{
1}{\log \left( \frac{W_2}{W_1} \right)}$ that ultimately is responsible for the success of the
new scaling relation.

Competitive growth models have the peculiar feature that at every deposition
one has to decide which of the two deposition rules is followed by the newly added particle.
We propose in the following a deposition model with similar features as the competitive growth models,
but where the competition is intrinsic and governed by the value of the substrate temperature.
This is a much more realistic situation, especially since in the growth
of thin films and nanostructures the substrate temperature is an important parameter that shapes to a large
extend the morphology of growing structures \cite{Jen99}.

The deposition model discussed in the following 
is based on Family's original RDSR process \cite{1986FF} and differs from this
model by the diffusion step. In the RDSR process a particle deposited on the surface is allowed to jump to one of the neighboring
sites if this site has a lower height than the site of deposition. In our model we assign an energy $E({\bf x},t) = f\, h({\bf x},t)$
to the column $\bf{x}$ where $h({\bf x},t)$ is the height of that column at time $t$. The constant $f$ can be thought
to be the gravitation constant.
Starting from an initially flat substrate, particles are deposited
on randomly chosen sites and then allowed to diffuse locally after deposition.
For a diffusion step taking place at time $t$, we select one of the neighboring
sites $\bf{y}$ at random and accept the jump with the temperature and time dependent (Metropolis like) probability
\begin{equation} \label{jump_rule}
P_{{\bf x} \longrightarrow {\bf y}}(T,t) = \left\{ \begin{array}{l}
1 ~,~~ \mbox{if} ~~ E({\bf y},t) 
\leq E({\bf x},t) \nonumber \\
e^{ - \left[ E({\bf y},t) - E({\bf x},t) \right]/k_B T} \\ 
= e^{- f \left[ h({\bf y},t) - h({\bf x},t)
\right]/k_B T }~,~~ \mbox{otherwise}. \nonumber
\end{array} ~.
\right.
\end{equation}
In the following we choose units thus that $f/k_B = 1$ where $k_B$ is the Boltzmann constant.

In contrast to the original RDSR model, we have in the present model a non-vanishing probability that a deposited
particle jumps to a neighboring site with a higher height than the deposition site.
We assume this jump to be thermally activated and to depend on the temperature $T$ of the substrate.
As we discuss elsewhere \cite{Chou09}, the substrate temperature is a parameter that allows the study
of novel properties of growing interfaces.

\begin{figure}[h]
\centerline{\epsfxsize=3.25in\ \epsfbox{
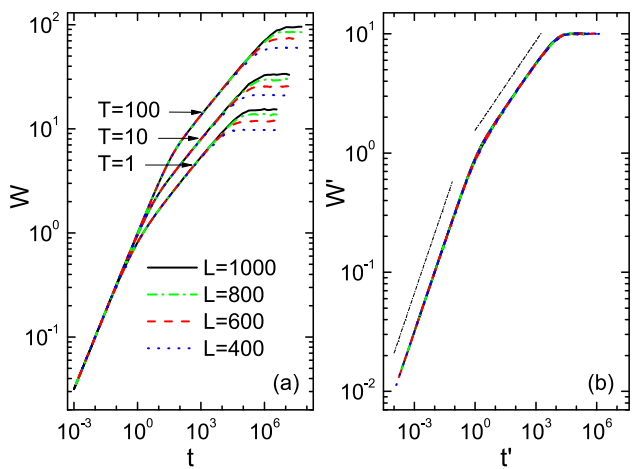}}
\caption{(a) Log-log plot of the surface width vs time for the temperature dependent deposition
model. Systems of different sizes at different temperatures are shown.
(b) A complete data collapse of all data sets is also achieved for this model.
The dashed lines indicate the expected
slopes in the RD and EW regimes.
\label{fig4}
}
\end{figure}

In Fig.\ \ref{fig4}a we show the temporal evolution of the width for various temperatures and system 
sizes. As for the RDSR process one distinguishes three regimes separated by two crossover points:
a RD regime, followed by a EW regime, with a final crossover to the saturation regime. In contrast to
the RDSR process, the random deposition process is not confined to the very early time regime $t \leq 1$ but can extend
to larger times. In fact, the crossover time $t_1$ between the random deposition and the EW regimes
is shifted to higher values for increasing temperatures and diverges in the limit of infinite temperatures.
Of special interest is that the surface widths shown in Fig.\ \ref{fig4}a
can be directly obtained from the Edwards-Wilkinson equation (\ref{eqEW}) for a system
of size $L$ with a temperature dependent diffusion constant. For example, for the temperatures shown
in Fig.\ \ref{fig4}a, the values of $\nu$ are: $\nu(T=100)=0.005$, $\nu(T=10) = 0.04$, and $\nu(T=1)
= 0.18$. 

We first check in Fig.\ \ref{fig4}b that also for the present model the scaling relation
(\ref{eqscal}) yields the full data collapse. Due to the simplicity of the model, we can
obtain the full information on the location of the two crossover points \cite{Chou09}.
In this way we find that, as usual, only the crossover to the saturation regime depends on the system
size. In addition, the coordinates of both crossover points display a linear dependence on the substrate
temperature. Taking these observations into account,
we can rewrite the scaling
relation (\ref{eqscal}) in the form
\begin{equation} \label{eqscal3}
W^\lambda = \left( a + b T \right)^{\lambda/2} G\left[ \frac{t^\lambda}{\left( a + b T \right)^\lambda}
\right]
\end{equation}
where $a=0.59$ and $b=0.29$,
with $\lambda(L) = 1/\mbox{log} \left( c \, L^\alpha \right)$ and $c = 0.55$, whereas
$\alpha = 1/2$ is the roughnes exponent of the EW universality class.
Eq. (\ref{eqscal3}) directly reveals for our model the dependence of the generalized scaling relation on
the system size and on the temperature. 

In conclusion, we have presented in this letter a parameter free scaling relation that yields
a complete data collapse for large classes of nonequilibrium growth processes with two crossover points. Examples include all simple
growth processes as well as more complex growing interfaces as encountered for example in competitive growth systems.
A deposition model with temperature dependent diffusion allows us to discuss the dependence
of the scaling relation on the relevant system parameters.


\begin{thebibliography}{}
\bibitem{Mea93} P.\ Meakin, Phys.\ Rep.\ {\bf 235}, 189 (1993).
\bibitem{Hal95} T.\  Halpin-Healy and Y.-C. Zhang, Phys.\ Rep.\ {\bf 254}, 215 (1995).
\bibitem{Bar95} A.-L. Bar\'{a}basi and H.\ E. Stanley, {\it Fractal Concepts in Surface Growth}
(Cambridge University Press, Cambridge, 1995).
\bibitem{Kru97} J. Krug, Advances in Physics {\bf 46}, 139 (1997).
\bibitem{Kru95} J. Krug, in {\it Scale Invariance, Interfaces, and Non-Equilibrium
Dynamics}, edited by A. McKane {\it et al.} (Plenum, New York, 1995), p. 25.
\bibitem{1982SE} S. F. Edwards and D. R. Wilkinson, Proc. R. Soc. Lond. A {\bf 381}, 17 (1982).
\bibitem{1986FF} F. Family, J. Phys. A {\bf 19}, L441 (1986).
\bibitem{1985FF} F. Family and T. Vicsek, J. Phys. A {\bf 18}, L75 (1985).
\bibitem{Kar86} M. Kardar, G. Parisi, and Y. Zhang, Phys. Rev. Lett. {\bf 56}, 889 (1986).
\bibitem{Wol90} D. E. Wolf and J. Villain, Europhys.\ Lett.\ {\bf 13}, 389 (1990).
\bibitem{Sar91} S. Das Sarma and P. Tamborenea, Phys. Rev. Lett. {\bf 66}, 325 (1991).
\bibitem{Hor01} C. M. Horowitz, R. A. Monetti, and E. V. Albano, Phys. Rev. E {\bf 63}, 066132 (2001).
\bibitem{Hor01a} C. M. Horowitz and E. V. Albano, J. Phys. A {\bf 34}, 357 (2001).
\bibitem{Cha02} A. Chame and F. D. A. Aar$\tilde{a}$o Reis, Phys. Rev. E {\bf 66}, 051104 (2002).
\bibitem{Hor03} C. M. Horowitz and E. V. Albano, Eur. Phys. J. B {\bf 31}, 563 (2003).
\bibitem{Kol04} A. Kolakowska, M. A. Novotny, and P. S. Verma, Phys. Rev. E {\bf 70}, 051602 (2004).
\bibitem{Mur04} D. Muraca, L. A. Braunstein, and R. C. Buceta, Phys. Rev. E {\bf 69}, 065103(R) (2004).
\bibitem{Bra05} L. A. Braunstein and C.-H. Lam, Phys. Rev. E {\bf 72}, 026128 (2005).
\bibitem{Iru05} I. Irurzun, C. M. Horowitz, and E. V. Albano, Phys. Rev. E {\bf 72}, 036116 (2005).
\bibitem{Hor06} C. M. Horowitz and E. V. Albano, Phys. Rev. E {\bf 73}, 031111 (2006).
\bibitem{Rei06} F. D. A. Aar$\tilde{a}$o Reis, Phys. Rev. E {\bf 73}, 021605 (2006).
\bibitem{Oli06} T. J. Oliveira, K. Dechoum, J. A. Redinz, and F. D. A. Aar$\tilde{a}$o Reis, Phys. Rev. E {\bf 74}, 
011604 (2006).
\bibitem{Kol06} A. Kolakowska, M. A. Novotny, and P. S. Verma, Phys. Rev. E {\bf 73}, 011603 (2006).
\bibitem{Kim89} J. M. Kim and J. M. Kosterlitz, Phys. Rev. Lett. {\bf 62}, 2289 (1989).
\bibitem{Chi04} C.-C. Chien, N.-N. Pang, and W.-J. Tzeng, Phys. Rev. E {\bf 70}, 021602 (2004).
\bibitem{Jen99} P. Jensen, Rev. Mod. Phys. {\bf 71}, 1695 (1999).
\bibitem{Chou09} Y.-L. Chou, M. Pleimling, and R. K. P. Zia, in preparation.
\end{thebibliography}
\end{document}